\documentclass[twocolumn,showpacs,preprintnumbers,amsmath,amssymb]{revtex4}
\usepackage{graphics}

\begin{document}

\title{Equal-Spin Pairing State of Superfluid $^3$He in Aerogel}

\author{Kazushi Aoyama and Ryusuke Ikeda}

\affiliation{%
Department of Physics, Graduate School of Science, Kyoto University, Kyoto 606-8502, Japan
}

\date{\today}
\begin{abstract}
The equal-spin pairing (ESP) state, the so-called A-like phase, of superfluid $^3$He in aerogels is studied theoretically in the Ginzburg-Landau (GL) region by examining thermodynamics, and the resulting equilibrium phase diagram is mapped out. We find that the ABM pairing state with presumably quasi long-ranged superfluid order has a lower free energy than the planar and "robust" states and is the best candidate of the A-like phase with a strange lowering of the polycritical point (PCP) observed experimentally. 
\end{abstract}



\maketitle

It is an important subject on liquid $^3$He in aerogel \cite{HS,Osheroff} to understand whether or not the pairing symmetry and other details of superfluid states are changed due to impurity effects provided by the aerogel background. At present, it will be clear that the B-like phase is, just as in bulk $^3$He, the BW pairing state with superfluid long range order (LRO), while it is a matter under debate whether or not the equal-spin pairing (ESP) state, the so-called A-like phase, near the superfluid transition temperature $T_c(P)$ is the 
ABM state. 

By assuming that the orientation of the ${\bf l}$-vector in the ABM pairing state should lose its LRO due to the impurity disorder, Volovik \cite{Volovik} has speculated that the A-like phase in zero magnetic field should be a glass state with the ABM pairing but with no superfluid rigidity (i.e., $\rho_s(q=0)=0$). Fomin \cite{Fomin} has argued with no free energy calculation that another ESP state, "robust" state, should be the A-like phase in aerogel. Among experimental suggestions on the A-like state, the {\it lowering} \cite{HS,Osheroff,Parpia} of the polycritical pressure (PCP) at which $T_c(P)$ and the A-B transition line $T_{\rm AB}(P)$ merge is a key observation: Although the range $T_c - T_{\rm AB}$ of the equilibrium A-like state at a fixed pressure $P$ above the bulk PCP is narrower in aerogels with lower porosity, the pressure range over which this state is detected seems to be {\it wider} with decreasing the porosity \cite{Osheroff,Parpia}. 
In this paper, we examine candidates of the A-like phase in equilibrium by calculating the superfluid free energy. We find that, due primarily to effects of quenched disorder on the {\it amplitude} of the pair-field (superfluid order parameter) $A_{\mu,i}$, the ABM state is the best candidate of the A-like state with the above-mentioned properties. 

To derive the Ginzburg-Landau (GL) free energy functional microscopically, we start from the quasiparticle hamiltonian with an effective interaction mediated by spin fluctuations \cite{BSA} and an impurity term 
${\cal H}_{\rm imp} = \int d^3r \sum_\sigma \psi^\dagger_\sigma({\bf r}) \, u({\bf r}) \, \psi_\sigma({\bf r})$, where $u({\bf r})$ is an impurity potential with zero mean. For a moment, impurity scattering effects inducing inhomogenuities of $A_{\mu,i}$ will be neglected. If the impurity effects on the quasiparticles are included just through their relaxation time, the GL functional is, as usual, expressed \cite{Th} by assuming a $p$-wave pairing interaction as the sum of ${\cal H}_{0}$ and ${\cal H}_{\rm gr}$, where 
\begin{eqnarray}\label{eq:GLMF}
{\cal H}_{0} [A_{\rho,k}] &=& \int_{\bf r} \! ( \alpha A_{\mu,i}^* A_{\mu,i} + \beta_1 |A_{\mu,i} A_{\mu,i}|^2 + \beta_2 (A_{\mu,i}^* A_{\mu,i})^2 \nonumber \\
&+& \beta_3 A^*_{\mu,i} A^*_{\nu,i} A_{\mu,j} A_{\nu,j} 
+ \beta_4 A^*_{\mu,i} A_{\nu,i} A^*_{\nu,j} A_{\mu,j} \nonumber \\
&+& \beta_5 A^*_{\mu,i} A_{\nu,i} A^*_{\mu,j} A_{\nu,j} ), 
\end{eqnarray} 
and 
\begin{equation}\label{eq:GLgrad}
{\cal H}_{\rm gr} [A_{\rho,k}] = \int_{\bf r} \! A^*_{\mu,i} \biggl( - K_1 \, \partial_i \partial_j A_{\mu,j} - \frac{K_2}{2} \, \nabla^2 A_{\mu,i} \biggr). 
\end{equation} 
Here, $\int_{\bf r} \! = \! \int \! d^3r$, 
$\alpha \! = \! N(0) [ \, {\rm ln}(T/T_{c0}) \! + \! \psi(1/2 \! + \! (4 \pi T \tau)^{-1}) - \psi(1/2) \, ]/3$, $K_1 \! = \! K_2 \! = \! 2 N(0) \xi_0^2/5$, 
and $
\xi_0 \! = \! v_{\rm F} \, \sqrt{- \psi^{(2)}(1/2+ (4 \pi T \tau)^{-1})/6} \, /(4 \pi T)$. Further, $T_{c0}$ is the mean field transition temperature in bulk at a fixed $P$, and $\tau^{-1} = 2 \pi N(0) \langle \, {\overline {|u_{{\bf p} - {\bf p}'}|^2}} \,\, \rangle_{\rm ang}$ is a relaxation rate, where the overbar and $\langle \cdot\cdot\cdot \rangle_{\rm ang}$ denote the averages over the randomness and the angle ${\rm cos}^{-1}({\bf p}\cdot{{\bf p}'}/k_{\rm F}^2)$, respectively. The coefficients 
$\beta_j$ are expressed as $\beta^{({\rm wc})}_j \! + \! \delta \beta_j^{({\rm sc})}$, where $\beta^{({\rm wc})}_2 = \beta^{({\rm wc})}_3 = \beta^{({\rm wc})}_4 = - \beta^{({\rm wc})}_5 = -2 \beta^{({\rm wc})}_1= 2 \beta^{({\rm wc})}(T) = - \psi^{(2)}(1/2+(4 \pi T \tau)^{-1}) \, \beta_0(T)/(7 \zeta(3))$, and 
$\beta_0(T) \! = \! 7 \zeta(3) N(0)/(240 \pi^2 T^2)$. 
The strong coupling corrections $\delta \beta_j^{({\rm sc})}$ with impurity effects included were obtained, following Brinkman et al.\cite{BSA}, in the spin-fluctuation model and, within the relaxation time approximation, satisfies the relations 
$\delta \beta_3^{({\rm sc})} \! = \! (\delta \beta_2^{({\rm sc})} \! + 5 \delta \beta_1^{({\rm sc})})/6$, 
$\delta \beta_4^{({\rm sc})} \! = \! \delta \beta_3^{({\rm sc})} \! + 5 \delta \beta_1^{({\rm sc})}$, and 
$\delta \beta_5^{({\rm sc})} \! = \! 7 \delta \beta_1^{({\rm sc})}$, where 
\begin{eqnarray}
\delta \beta_1^{({\rm sc})} &=& - 3.3 \times 10^{-3} \, \beta_0(T) \, t \, \delta \, \sum_m (D_1^{(d)}(m))^2, \nonumber \\
\delta \beta_2^{({\rm sc})} &=& \delta \beta_1^{({\rm sc})} \sum_m [9 (D_2^{(d)}(m))^2 - 6 D_1^{(d)}(m) D_2^{(d)}(m) \nonumber \\
&-& 2(D^{(d)}_1(m))^2]/[\sum_m (D_1^{(d)}(m))^2], 
\end{eqnarray}
$D_1^{(d)}(m) = [\, |m|^{-1} + (|m| + (2 \pi \tau T)^{-1})^{-1} \,] \times [\psi(1/2 + |m| + (4 \pi T \tau)^{-1})- \psi(1/2+(4 \pi \tau T)^{-1})]/2$, $D_2^{(d)}(m)=\psi^{(1)}(1/2 + |m| + (4 \pi \tau T)^{-1})/2$, and the parameter $\delta$ was defined in eq.(3.10) of Ref.\cite{BSA}. We find that, due primarily to the prefactor $t=T/T_{c0}$ in $\delta \beta_j^{({\rm sc})}$, the strong-coupling corrections are weakened with increasing the impurity strength $(2 \pi T \tau)^{-1}$ in qualitative agreement with the results in a static approximation \cite{BK}. 

In the relaxation time approximation used in obtaining eqs.(\ref{eq:GLMF}) and (\ref{eq:GLgrad}), the impurity-ladder vertex corrections are neglected by assuming $2 \pi T \tau \gg 1$. Such vertex corrections are, at most, O($1/(2 \pi \tau T)$) in magnitude and do not change the resulting phase diagram qualitatively, because they appear only as slight changes of $\alpha$ and $\beta_0$, i.e.,  in a manner independent of the pairing states. 
There are also additional terms \cite{Th,BK} composed of a single impurity line in $\beta^{({\rm wc})}_m$ leading to the change of coefficients, $\beta^{({\rm wc})}_m \to \beta^{({\rm wc})}_m - (-1)^m \beta_0 \varepsilon_{\rm imp}$ ($m=2$, $4$, or $5$), where $\varepsilon_{\rm imp}$ is of the order $(2 \pi T \tau)^{-1}$. The reason for our neglect of $\varepsilon_{\rm imp}$-terms will be explained later. 

To describe an inhomogenuity of $A_{\mu,i}$ due to the impurity scatterings, an additional quadratic term 
\begin{eqnarray}\label{eq:GLpin} 
{\cal H}_{2,{\rm d}} &=& T \sum_{q,q'} A_{\mu,i}^*({\bf q}) A_{\mu,j}({\bf q}') \sum_\varepsilon \biggl[ \frac{\delta_{i,j} \delta_{q,q'}}{3} \frac{2 \pi N(0)}{2|\varepsilon| + \tau^{-1}} \nonumber \\
&-& \int \frac{d^3p d^3p'}{(2 \pi)^6} {\hat p}_i {\hat p}'_j {\cal G}_{\varepsilon} ({\bf p}+{\bf q}/2, {\bf p}'+{\bf q}'/2) \nonumber \\
&\times& {\cal G}_{-\varepsilon} (-{\bf p}+{\bf q}/2, -{\bf p}'+{\bf q}'/2) \biggr],
\end{eqnarray}
usually neglected in the mean field (MF) analysis, needs to be evaluated, where ${\cal G}_\varepsilon ({\bf p}, {\bf p}')$ is the quasiparticle propagator defined prior to the random average and with a Matsubara frequency $\varepsilon$. By expanding eq.(\ref{eq:GLpin}) in powers of $u({\bf r})$, its O($u({\bf r})$) term identically vanishes, and its O($[u({\bf r})]^2$) term is evaluated as 
\begin{eqnarray}\label{eq:GLpin1}
{\cal H}_{2,{\rm d}} &\simeq& T \sum_{q,q'} A_{\mu,i}^*({\bf q}) A_{\mu,j}({\bf q}') \sum_\varepsilon \frac{\pi^2 k_{\rm F} N(0)}{2 E_{\rm F} \, (2 |\varepsilon| + \tau^{-1})^{2}} \nonumber \\
&\times& \int \frac{d^3k}{(2 \pi)^3} \frac{k_i k_j}{|k|^3} u_{{\bf k}+{\bf q}} \, u_{-{\bf k} - {\bf q}'}, 
\end{eqnarray}
where we have assumed the ${\bf k}$-integral to be dominated by large $|{\bf k}|$ values (see the $\gamma$'s expression below). 
An additional quartic term arising from eq.(\ref{eq:GLpin1}) after the random 
average becomes 
\begin{eqnarray}\label{eq:rep4}
{\cal H}_{\rm d}^{(n)} &\simeq& - \frac{3}{10} \, \delta \beta_{\rm d} \sum_{a,b=1}^n \sum_{q_j} 
\! (\delta_{i,j} \delta_{r,s} + \delta_{i,r} \delta_{j,s} + \delta_{i,s} \delta_{r,j} ) \nonumber \\
&\times& \delta_{\sum q, 0} (A^{(a)}_{\mu,i}(q_1) A^{(b)}_{\nu,r}(q_3))^* A^{(a)}_{\mu,j}(q_2) A^{(b)}_{\nu,s}(q_4), 
\end{eqnarray}
where $\sum q \! = \! q_1 \! + \! q_3 \! -\! q_2 \! - \! q_4$, $a$ and $b$ are replica indices
, 
\begin{equation}\label{eq:dbeta}
\delta\beta_{\rm d} = \beta_0 \frac{\gamma}{E_{\rm F} T \tau^2} \frac{5 \pi^4}{42 \zeta(3)} \biggl( \psi^{(1)}\bigl(\frac{1}{2}+\frac{1}{4 \pi T \tau} \bigr) \biggr)^2, 
\end{equation}
and $
\gamma \equiv (\tau N(0))^2 \int dk \, \theta(k_{\rm F} - k) \, \theta(k) \, {\overline {|u_{\bf k}|^4}}/(2 \pi^2 k_{\rm F})$. 

We suppose that scattering events in an aerogel have local and strong {\it anisotropies} \cite{HS}, and that such {\it random} anisotropies leading to a pinning of the ${\bf l}$-vector in the ABM state be described as a strong ${\hat k}$-dependence of $|u_{\bf k}|^2$. In fact, if $|u_{\bf k}|^2$ is a function only of $k^2$, and $|q|$, $|q'| \ll k_{\rm F}$, eq.(\ref{eq:GLpin1}) becomes an isotropic term like $\sim \int_{\bf r} A^*_{\mu,i} A_{\mu,i}$ independent of the pairing state. 
Below, a strong random anisotropy such that ${\overline {|u_{\bf k}|^4}} \gg ( \, {\overline {|u_{\bf k}|^2}} \, )^2$ is assumed. Further, since higher order terms in $u({\bf r})$ neglected above will also enhance $\delta \beta_{\rm d}$, the {\it effective} value of the parameter $\gamma$ may be large, e.g., of order unity, in $^3$He in an aerogel. Then, $\delta \beta_{\rm d}$ can become comparable with $|\delta \beta_j^{({\rm sc})}|$. Under these assumptions, the replicated GL functional ${\cal H}_{\rm GL}^{(n)}$ is given by ${\cal H}_{\rm GL}^{(n)} \! = \! \sum_a ({\cal H}_{0} [A_{\mu,i}^{(a)}] + {\cal H}_{\rm gr} [A_{\mu,i}^{(a)}] ) + {\cal H}_{\rm d}^{(n)}$. It is convenient to rewrite ${\cal H}_{\rm d}^{(n)}$ in the ABM state as ${\cal H}_{{\rm d} (1)}^{(n)} + {\cal H}_{{\rm d} (2)}^{(n)}$, where 
\begin{eqnarray}\label{eq:rep1}
{\cal H}_{{\rm d} (1)}^{(n)} &=& - \frac{3}{5} \, \delta\beta_{\rm d} \sum_{a,b} \int_{\bf r} |\Delta^{(a)}({\bf r}) \Delta^{(b)}({\bf r})|^2, 
\end{eqnarray}
and 
\begin{eqnarray}\label{eq:rep2}
{\cal H}_{{\rm d} (2)}^{(n)} &=& \frac{3}{20} \delta\beta_{\rm d} \sum_{a,b} \int_{\bf r} |\Delta^{(a)}({\bf r}) \Delta^{(b)}({\bf r})|^2 \nonumber \\
&\times& [ \, 1 - ({\bf l}^{(a)}\cdot{\bf l}^{(b)})^2 \, ]. 
\end{eqnarray} 
For the BW and ESP states, ${\cal H}_{{\rm d} (2)}$ does not appear, and the factor $3/5$ in eq.(8) is replaced by $1/2$. For a moment, we focus on the ABM pairing state. 

Below, the free energy will be examined in terms of the Gaussian variational method (GVM) \cite{GD,Dotsenko} which is often used for random systems. By assuming a trial Gaussian functional ${\cal H}_g^{(n)}$, the free energy is approximated in GVM by 
\begin{equation}\label{eq:free energy}
F = F_g + \frac{1}{n} \langle {\cal H}_{\rm GL}^{(n)} - {\cal H}_g^{(n)} \rangle_g, 
\end{equation}
where $F_g$ is the free energy for ${\cal H}_g^{(n)}$ divided by $n$, $\langle \, \, \, \rangle_g$ is the ensemble average on ${\cal H}_g^{(n)}$, and the $n \to 0$ limit is taken at the end. It is reasonable to assume ${\cal H}_g^{(n)}$ to take the form ${\cal H}_{g,{\rm amp}} + {\cal H}_{g,{\rm sym}}$, where ${\cal H}_{g,{\rm amp}}$ (${\cal H}_{g,{\rm sym}}$) is a Gaussian functional composed {\it only} of $|\Delta|$-fluctuations (of symmetry variables such as the phase and ${\hat l}$-vector). Then, the fact that ${\cal H}_{\rm GL}$ appears in eq.(\ref{eq:free energy}) {\it only} as its average permits us to use eq.(\ref{eq:free energy}) after averaging over ${\bf {\hat l}}$-orientations in terms including $\partial |\Delta|$ in $\sum_a {\cal H}_{\rm gr}[A_{\mu,i}^{(a)}]$. It can be accomplished with no knowledge on ${\cal H}_{g,{\rm sym}}$ in the present context with no strict LRO of the ${\bf {\hat l}}$-orientation, i.e., ${\overline {\, {\hat {\bf l}} \, }}=0$. Consequently, ${\cal H}_{\rm GL}$ can be replaced by the sum of the the amplitude part ${\cal H}_{\rm amp}^{(n)}= {\tilde {\cal H}}_{\rm amp}^{(n)}+{\cal H}_{{\rm d}(1)}^{(n)}$, where 
\begin{equation}\label{eq:amp}
{\tilde {\cal H}}_{\rm amp}^{(n)} = \sum_{a=1}^n \int_{\bf r} (\alpha |\Delta^{(a)}|^2 + \frac{5 K_1}{6} (\nabla |\Delta^{(a)}|)^2 + \beta_{\rm A} \, |\Delta^{(a)}|^4 ), 
\end{equation}
and the symmetry variables' part ${\cal H}_{\rm sym}^{(n)} = {\tilde {\cal H}}_{\rm gr}^{(n)} + {\cal H}_{{\rm d}(2)}^{(n)}$, where  
\begin{equation}\label{eq:sym}
{\tilde {\cal H}}_{\rm gr}^{(n)} = \frac{-K_1}{2} \sum_{a=1}^n \int_{\bf r} |\Delta|^2 (a_{\mu,j}^{(a)})^* ( 2 \partial_i \partial_j a_{\mu,i}^{(a)} 
\! + \! \nabla^2 a_{\mu,j}^{(a)}), 
\end{equation} 
$\beta_{\rm A} \equiv \beta_{245}$, $\beta_{ijk}=\beta_i+\beta_j+\beta_k$, and $A_{\mu,i}=|\Delta| 
a_{\mu,i}$. 
The total superfluid free energy becomes the sum of $F_{\rm amp}$ and $F_{\rm sym}$, which are obtained below from ${\cal H}_{\rm amp}^{(n)}$ and ${\cal H}_{\rm sym}^{(n)}$, respectively. 

Results on $F_{\rm amp}$ are obtained by following the GVM analysis for the GL model of the Ising spin system \cite{Dotsenko}. Since we focus on the GL region far below the critical region, the relations $(T \beta_{\rm A})^2/[(N(0))^4 \xi_0^6] \ll (T_c - T)/T_{c0} < 1$ can be assumed together with $\delta \beta_{\rm d} \ll \beta_0$. Then, following the analysis in Ref.\cite{Dotsenko} closely, $F_{\rm amp}$ 
is expressed by 
\begin{equation}\label{eq:freeamp}
\frac{F_{\rm amp}}{V} \simeq - \, \frac{N(0) \, |\lambda_p| \, |\Delta_{\rm MF}|^2}{2} - {\rm O}(T/\xi_0^3), 
\end{equation}
where $V$ is the volume, the ${\rm O}(T/\xi_0^3)$ term is {\it independent of} the pairing symmetry, 
\begin{equation}\label{eq:tc}
\lambda_p = \frac{\alpha}{N(0)} + \frac{3^{5/2} T}{(N(0))^2 \xi_0^3} \, \frac{\beta_{\rm A} - 2 \delta \beta_{\rm d}/5}{\pi} 
\end{equation}
corresponding near $T_c$ to $(T - T_c)/T_{c0}$, and $
|\Delta_{\rm MF}|^2 = N(0) |\lambda_p|/(2 \beta_{\rm A})$. 
Note that $T_c$ depends on the pairing state through the terms in eq.(\ref{eq:tc}) except $\alpha/N(0)$. In fact, in clean limit with no $\delta \beta_{\rm d}$, the relation $\beta_{\rm A}=\beta_{\rm B} \equiv \beta_{12}+\beta_{345}/3$ defining the bulk PCP easily follows under the condition that a state with higher $T_c$ is realized just below the resulting $T_c(P)$. Further, other fluctuation corrections of ${\rm O}(T|\lambda_p|/\xi_0^3)$ were dropped in $F_{\rm amp}$ because, in the free energy {\it difference} between the ABM and BW states, the contribution from these corrections is of higher order in $T/E_{\rm F}$ compared with that from $F_{\rm sym}$ (see eq.(\ref{eq:freesym}) below). That is, $F_{\rm amp}$ is well approximated by the condensation energy with $\alpha$ replaced by $N(0) \lambda_p$. 

Hereafter, eq.(\ref{eq:sym}) will be treated in the London limit by replacing $|\Delta|$ there by $|\Delta_{\rm MF}|$. 
Just as in $F_{\rm amp}$, the $\delta \beta_{\rm d}$-independent fluctuation contribution to $F_{\rm sym}$ will be neglected, and we focus below on the difference $\Delta F_{\rm sym} \! = \! F_{\rm sym}(\delta \beta_{\rm d}) \! - \! F_{\rm sym}(0)$. 
However, a direct application of GVM to ${\cal H}_{\rm sym}^{(n)}$ 
is not easy due to the 3D nature of ${\bf {\hat l}}$. For this reason, the simpler model 
\begin{equation}\label{eq:iso}
{\cal H}_{{\rm sym} (0)}^{(n)} = \frac{|\Delta_{\rm MF}|^2}{2} \sum_a \int_{\bf r}  K_t \, \partial_i {\hat l}_j^{(a)} \partial_i {\hat l}_j^{(a)} + {\cal H}_{{\rm d}(2)}^{(n)} 
\end{equation}
will be first considered by assuming ${\bf {\hat l}}$ to be a planar vector (${\bf {\hat l}} \! = \! {\hat x} \, {\rm cos}\beta \! + \! {\hat y}\, {\rm sin}\beta$). For the model eq.(\ref{eq:iso}), $\Delta F_{\rm sym}/V$ in GVM was examined elsewhere \cite{GD} to derive the quasi LRO and, when $|\Delta_{\rm MF}|^2 \delta \beta_{\rm d} \, \xi_0^2 /K_t$ is small enough, is given by $- 3 T q_c |\lambda_p| \delta\beta_{\rm d} N(0)/(40 \pi^2 K_t \beta_{\rm A}) < 0$, where $q_c$ is a momentum cutoff of the order $\xi_0^{-1}$. This result is {\it insensitive} to the quasi LRO of ${\bf {\hat l}}$-orientation in the model (\ref{eq:iso}) and is the same as the $T \to 0$ limit \cite{Larkin} of the corresponding random-force model, which is obtained after the replacement, $({\bf {\hat l}}^{(a)}\cdot{\bf {\hat l}}^{(b)})^2 \to 1 - (\beta^{(a)}-\beta^{(b)})^2$, in ${\cal H}_{{\rm d}(2)}^{(n)}$ in eq.(\ref{eq:iso}). Based on this fact, we will assume that, in order to evaluate the thermodynamic quantities, ${\cal H}_{\rm sym}^{(n)}$ can also be replaced by the corresponding 
random-force model. Then, after a lengthy but straightforward calculation, 
we obtain 
\begin{equation}\label{eq:freesym}
\frac{\Delta F_{\rm sym}(\delta \beta_{\rm dis})}{V} \simeq \frac{ - 9}{20 \pi} \frac{T \, |\lambda_p|}{\beta_{\rm A} \, \xi_0^3} \delta \beta_{\rm d}. 
\end{equation}
For the present purpose, the sum of eqs.(\ref{eq:freeamp}) and (\ref{eq:freesym}) can be regarded as the total free energy in the ABM state. 

Here, the above results on the ABM state will be compared with those on other pairing states. The {\it total} free energy and $\lambda_p$ in the BW ("robust" ESP) state are given by eq.(\ref{eq:freeamp}) and eq.(\ref{eq:tc}) if replacing $\beta_{\rm A}$ and $2 \delta \beta_{\rm d}/5$ there by $\beta_{\rm B}$ ($\beta_{\rm R} \equiv \beta_2 + (\beta_{13}+5\beta_{45})/9$) and $\delta \beta_{\rm d}/3$, respectively, where $\beta_{ij}=\beta_i+\beta_j$, while $F_{\rm amp}$ and $\lambda_p$ for the planar state are given simply by replacing $\beta_{\rm A}$ in them by $\beta_{\rm P} \equiv \beta_{12}+\beta_{345}/2$. Since it is easily seen that $\beta_{\rm R} > \beta_{\rm P} > \beta_{\rm A}$ for any $\tau T_{c0}$ value we have examined, higher $T_c(P)$-values are obtained only for the ABM or BW states, and hence, the pairing states just below $T_c(P)$-curve are the ABM or BW states. This result just below $T_c$ is unaffected by including the $\varepsilon_{\rm imp}$-terms in $\beta_m$ (see the paragraph prior to eq.(\ref{eq:GLpin})). In fact, $\beta_{\rm N}$ is replaced then by $\beta_{\rm N} - \beta_0 \varepsilon_{\rm imp}$ (${\rm N} =$ A, B, P, or R) {\it commonly} for the four pairing states. Further, the "robust" \cite{Fomin} and planar states cannot occur even as an intermediate state between the ABM and BW states. This conclusion on the "robust" state with no $\Delta F_{\rm sym}$ is clear at this stage. We have also evaluated $ - \Delta F_{\rm sym}$ for the planar state and found that it is at most $T |\lambda_p| \delta \beta_{\rm d}/(10 \pi \beta_{\rm P} \xi_0^3)$, which cannot cancel out the cost in $F_{\rm amp}$ relative to the ABM state. 

Examples of computed $T_c(P)$ and $T_{\rm AB}(P)$ curves are shown in Fig.1. The $T_c(P)$-curve is defined from a larger value of $T_c$ for each of the ABM and BW states, while $T_{\rm AB}(P)$ is obtained simply by comparing the free energies of the two states with each other. In our computation, we have assumed $\varepsilon_{\rm imp}=0$. We have numerically verified that no visible change of $T_{\rm AB}$ occurs by choosing $\varepsilon_{\rm imp} = \pm (2 \pi \tau T)^{-1}$. We have used $E_{\rm F}(P)$ data in Ref.\cite{Kuroda} and have assumed $\delta = 300 \, T_{c0}/E_{\rm F}(P)$ together with a typical $T_{c0}(P)$-curve of bulk $^3$He. The curve with $(2 \pi \tau)^{-1}=0.18$ (mK) well explains why the A-like region was not easily observed on warming \cite{Parpia} in spite of a significantly low PCP. Over a wide range of $\tau^{-1}$-values, with increasing $\tau^{-1}$, 
the width $T_c(P) - T_{\rm AB}(P)$ shrinks, while PCP tends to be {\it lowered} compared to the bulk PCP. The former feature is a reflection of a decrease of the strong-coupling correction to $\beta_{\rm N}$, while the latter is due to an increase of $\delta\beta_{\rm d}$. For low enough $\tau^{-1}$ values (the $(2 \pi \tau)^{-1}=0.13$ curve in Fig.1), however, PCP is slightly higher than the bulk PCP. Such a nonmonotonic $\tau$-dependence of PCP is due to a competition between the two impurity-induced effects 
mentioned above. On the other hand, a reduction of $\gamma$ (random anisotropy) under a fixed $\tau$ monotonously shrinks $T_c - T_{\rm AB}$ and increases PCP. It is possible that a decrease of the porosity in aerogel corresponds to increases not only of $\tau^{-1}$ but also of $\gamma$. Then, the PCP in real aerogels may be monotonously lowered with decreasing the porosity. 

\begin{figure}[t]
\scalebox{0.75}[0.45]{\includegraphics{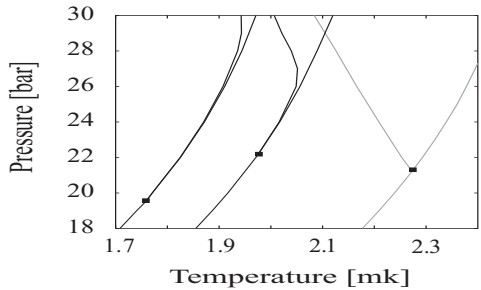}}
\caption{$T_c(P)$ and $T_{\rm AB}(P)$ curves for $(2 \pi \tau)^{-1} = 0$ (mK) (dotted curves), $0.13$ 
(thick solid), and $0.18$ 
(thin solid). Each PCP is denoted as a solid square on each $T_c(P)$-curve. For all curves, $\gamma=4.6$ was commonly used. } 
\label{fig:3he} \end{figure}

Finally, we comment on the ordering at long distances in the ABM state in aerogel in zero magnetic field by assuming a weak disorder at which no {\it singular} topological defects appear. Let us first consider the isotropized gradient energy ${\cal H}'_{{\rm gr} {\rm A}} + {\cal H}_{{\rm sym}(0)}^{(n)}$, where 
\begin{equation}\label{eq:gradd} 
{\cal H}'_{{\rm gr} {\rm A}} = \frac{1}{2} \int_{\bf r} \biggl[ {\tilde \rho}_s  ({\bf v}_s^{\rm T})^2 + \lambda_s ({\rm div}{\bf l})^2 + \lambda_b (({\bf l}\cdot\nabla){\bf l})^2 \biggr], 
\end{equation}
in place of ${\cal H}_{\rm sym}^{(n)}$ for the ABM state. Here, $\lambda_s$, $\lambda_b > -|\Delta_{\rm MF}|^2 K_t$, and ${\bf v}_s^{\rm T}$ is the transverse component of velocity leading to {\it nonsingular} vortices \cite{MH} which might affect the ordering. In the limiting case with no first term, corresponding 
to the dipole-locked case in nonzero magnetic field where ${\bf {\hat l}}$ is a planar vector (see eq.(\ref{eq:iso})), the quasi LRO of ${\bf l}$-orientation in the above model is already known in the context of nematic glass \cite{Feldman}. To examine effects of {\it nonsingular} vortices on this quasi LRO, the $({\bf v}_s^{\rm T})^2$ term will be expressed using the Mermin-Ho relation \cite{MH} as $\int_{\bf r} \int_{{\bf r}'} \, {\bf \Omega}({\bf r}) \cdot {\bf \Omega}({\bf r}')/(4 \pi |{\bf r} - {\bf r}'|)$, where ${\bf \Omega} \propto \varepsilon_{ijk} l_i (\nabla l_j \times \nabla l_k)$. The spatial nonlocality implies that this term is unrenormalized. Further, in the perturbative renormalization analysis \cite{Feldman} under a fixed ${\bf l}^2$, this term quadratic in ${\bf \Omega}$ is quartic in the "fast" variables like $|{\bf l} \times \delta {\bf l}|$. It implies that the $({\bf v}_s^{\rm T})^2$ term is an irrelevant perturbation to the $T=0$ fixed point controlling the ${\bf l}$'s quasi LRO. For the same reason, other ${\bf v}_s^{\rm T}$-related terms, neglected in eq.(\ref{eq:gradd}), of ${\cal H}_{\rm sym}^{(n)}$ are also irrelevant. Thus, the ${\bf {\hat l}}$'s quasi LRO is expected even in zero magnetic field. Since the pair-field $A_{\mu,i}$ is linear in ${\bf d}$ and the orbital "triad" including ${\bf l}$, the ${\bf l}$'s quasi LRO should result in the superfluid quasi LRO in any 
field \cite{com1}. 

In conclusion, the free energy calculation including effects of quenched disorder on the pair-field results in a phase diagram of $^3$He in aerogel qualitatively consistent with experimental ones \cite{Osheroff,Parpia} and shows that the ABM pairing state describes the A-like state of $^3$He in aerogel. Bearing the quasi LRO and a recent observation \cite{Halperin} of an A$_1$-A$_2$ transition in mind, it is believed that the A-like state in $^3$He in aerogel at weak disorder should be a glass phase with the ABM pairing and superfluid quasi LRO. 

We are grateful to I.A. Fomin, K. Nagai, and T. Mizusaki for discussions. 


\end{document}